%% file: channel-access-infocom14-v3.tex
\documentclass[conference]{IEEEtran}

\makeatletter
\def\ps@headings{%
\def\@oddhead{\mbox{}\scriptsize\rightmark \hfil \thepage}%
\def\@evenhead{\scriptsize\thepage \hfil \leftmark\mbox{}}%
\def\@oddfoot{}%
\def\@evenfoot{}}
\makeatother
\usepackage{color}
\makeatletter

\newtheorem{theorem}{Theorem}

\newcommand{\Rmnum}[1]{\expandafter\@slowromancap\romannumeral #1@}
\makeatother
\usepackage{verbatim}
\usepackage{cite}
\ifCLASSINFOpdf
   \usepackage[pdftex]{graphicx}
\else
 \usepackage[dvips]{graphicx}
\fi

\usepackage[cmex10]{amsmath}
\usepackage{latexsym}
\usepackage{amssymb}
\usepackage{pdfsync}
\usepackage{algorithm}
\usepackage{algorithmic}

\usepackage[dvips]{graphicx}
\usepackage[tight,footnotesize]{subfigure}
\usepackage{sidecap}

\usepackage{caption}

\usepackage{latexsym}
\usepackage{amssymb}
\usepackage{verbatim}
\usepackage{amsmath}

 \usepackage[dvips]{graphicx}
 \usepackage[tight,footnotesize]{subfigure}

\begin{document}
\input{title}
\maketitle
\input{abstract}

\section{Introduction}
\label{sec:introduction}
\input{introduction}

\vspace{-0.15in}
\section{Network model}
\label{sec:model}

\input{networkmodel}

\vspace{-0.15in}
\section{Problem formulation}
\label{sec:formulation}

\input{formulation}

\vspace{-0.15in}
\section{Channel access}
\label{sec:channelaccess}

\input{channelaccess}

    \subsection{The learning policy}
        \input{channelaccess-sub-learning}
    \subsection{Centralized approximation solution for channel access}

\input{channelaccess-sub-centralized}
    \subsection{Distributed channel access}
        \input{channelaccess-sub-distributed}

    \subsection{Improve to constant-time-complexity strategy decision}
        \input{channelaccess-sub-improve}

        \vspace{-0.15in}
    \subsection{Practical regret}
        \vspace{-0.05in}
        \input{channelaccess-sub-regret}

\section{Simulations}
\label{sec:simulation}

\input{simulation}

\vspace{-0.15in}
\section{Related works}
\label{sec:relatedwork}

\input{relatedwork}

\vspace{-0.15in}
\section{Conclusion}
\label{sec:conclusion}
\input{conclusion}

\bibliographystyle{IEEEtran}
\input{channel-access-infocom14-v3.bbl}



\end{document}

%% file: title.tex
\title{\gdef\thefootnote{}Almost Optimal Channel Access in Multi-Hop Networks With Unknown Channel Variables       }

\author{\IEEEauthorblockN{Yaqin Zhou\IEEEauthorrefmark{1}, Xiang-yang Li\IEEEauthorrefmark{2}\IEEEauthorrefmark{3}, Fan Li\IEEEauthorrefmark{4}, Min Liu\IEEEauthorrefmark{1}, Zhongcheng Li\IEEEauthorrefmark{1}, Zhiyuan Yin\IEEEauthorrefmark{4}}
\IEEEauthorblockA{\IEEEauthorrefmark{1}Institute of Computing Technology, Chinese Academy of Sciences, Beijing, China}
\IEEEauthorblockA{\IEEEauthorrefmark{2}Illinois Institute of Technology, Chicago, IL, USA}
\IEEEauthorblockA{\IEEEauthorrefmark{3}School of Software and TNLIST, Tsinghua University, China }
\IEEEauthorblockA{\IEEEauthorrefmark{4}School of Computer Science, Beijing Institute of Technology}} 

%% file: abstract.tex
\begin{abstract}
  We consider distributed channel access in multi-hop cognitive radio
  networks.
  Previous works on opportunistic channel access
  using multi-armed bandits (MAB)  mainly
  focus on single-hop networks that assume complete conflicts among
  all secondary users. In the multi-hop multi-channel network settings
  studied here, there is more general competition among different
  communication pairs.
We  formulate the problem as a linearly combinatorial MAB problem that
  involves a maximum weighted independent set  (MWIS)
  problem with unknown weights which need to learn.
Existing methods for MAB where each of $N$ nodes
  chooses from $M$  channels  have exponential time and space
  complexity $O(M^N)$, and poor theoretical guarantee on throughput
  performance.
We   propose a
  distributed channel access algorithm that can achieve $1/\rho$ of the
  optimum averaged throughput where each node has communication complexity $O(r^2+D)$
  and  space complexity $O(m)$ in the learning process, and  time
  complexity  $O(D m^{\rho^r})$ in strategy decision process for an
  arbitrary wireless network. Here $\rho=1+\epsilon$ is the
  approximation ratio to MWIS for a  local $r$-hop network with $m<N$ nodes,
  and $D$ is the number of mini-rounds inside each round of strategy decision.
  For randomly located networks with an average degree $d$, the time
  complexity is $O(d^{\rho^r})$.
\end{abstract}

%% file: introduction.tex
Available spectrum is being exhausted, while a lot of frequency bands are extremely under utilized. As a promising solution to improve dynamic allocation of the under-utilized spectrum, cognitive radio technology allows secondary users to opportunistically access vacant channels in temporal and spatial domain when the primary user is idle.
 However,
 due to resource and hardware constraints, at a given time, cognitive radios (CR) can sense only a part of heterogeneous channels with unknown quality before transmission.

 Thus, it is core for secondary users  to learn and select the  best possible channels to access.
Several recent results \cite{liu2010distributedmab}, \cite{tekin2012online}, \cite{kalathil2012decentralized}, \cite{liu2013restlessmab}, \cite{anandkumar2010opportunistic}, \cite{anandkumar2011distributed}, \cite{gai2011decentralized}
are proposed to take the dynamic spectrum sharing problem as the
multi-armed bandits problem, and attempt to find a  dynamic channel
access policy that results in almost optimal expected throughput (or
\emph{zero-regret}) through learning history, compared with the
optimal fixed channel policy. However, these methods generally adopt
the simplest form of MAB where only single-hop networks fit the
model. Dynamic channel access in multihop cognitive radio networks
demands more  sophisticated formulation that considers constraints of
general interference among users. A naive extension of formulation
from the single-hop case to multihop case will lead to regret, time
and space complexity that is exponential with the number of users in
the learning process. More specifically, taken as an arm a strategy
consisting of decisions from each of the $N$ users, there will be $O(M^N)$
combinations totally when each user has $M$ channels to choose. As all
these aforementioned works adopt a
UCB-type learning policy \cite{lai1985ucb} \cite{auer2002finite}
\cite{agrawal1995sample}, the upper bound of regret as well as time
and space complexity is linear with the number of arms, thus   linear
with $O(M^N)$ in multihop networks.

Efficient channel access under multihop networks also requires decentralized design with low computation and communication.
Previous decentralized MAB methods \cite{liu2013restlessmab} \cite{tekin2012online} \cite{gai2012mab} pay little attention to these practical challenges around multihop networks.
Though there is no communication cost in \cite{liu2013restlessmab}\cite{tekin2012online}, they require exponential time in a single learning round. Distinct from \cite{liu2013restlessmab}, \cite{tekin2012online} assumes multiple users can access the same resource, which does not capture conflicts among near-by users.
On the other hand, \cite{li2012almost} proposes a low-computation learning policy for  multi-hop networks, but the policy takes a centralized form and still leaves challenges on distributed implementation unsolved.

Here we investigate the problem of achieving maximum expected
throughput through a decentralized learning process with low
computation and communication cost. As this problem involves
competition  among adjacent  users, and cooperation
for maximum throughput network wide, there may be no effective
solutions if we directly formulate the problem into an integer linear
programming. We then subtly formulate the problem into a linearly
combinatorial MAB problem that shall find a maximum weighted
independent set of vertexes where weight is unknown channel quality.
This novel formulation facilitates us to utilize a zero-regret learning policy 
where it only costs time and space complexity $O(MN)$ for a network with $M$ channels 
and $N$ secondary users.
The other  benefit is that  we can adaptively choose  efficient
 methods to solve the involved NP-hard MWIS problem and
 still achieve zero-regret.

We   propose a decentralized channel access scheme based on robust
PTAS \cite{nieberg2005robust} to approximately solve the MWIS
problem. Our decentralized implementation achieves an approximation
ratio of $\rho=1+\epsilon$, but only requires  time complexity
$O(Dm^{\rho^r})$ to find the strategy decision after weight is
estimated.
Here $r=O(\log_{\rho} M)$ is the hop number required to achieve a
robust PTAS\cite{nieberg2005robust}, which is a constant for networks
with a constant number of channels to choose.
It costs time complexity $O(d^{\rho^r})$ for a random
network.
Our simulation results show that our new distributed learning policy
indeed outperforms previous policies in terms of average throughput,
time and storage cost.

The remainder of this paper is organized as follows. We present the
network model in Section \ref{sec:model},  problem formulation in
Section \ref{sec:formulation}, our distributed access policy in
Section~\ref{sec:channelaccess}, and our simulation results in
Section \ref{sec:simulation}.
We review related work in Section
\ref{sec:relatedwork}, and conclude the work in Section
\ref{sec:conclusion}.

%% file: networkmodel.tex
Consider a network  $G=(V,E,C)$ with  a set $V=\{v_i|i=1,\dots,N\}$ of $N$ nodes (users), a set $E$ of edges denoting conflicts, and a set $C=\{c_j|j=1,\dots,M\}$ of $M$ channels. We assume $M$ is a constant as the number of available frequency bands is fixed in a given network.
We use unit disks to model conflicts between two nodes, where each node is treated as a  disk centered on itself.
Conflicts happen if any two intersected disks access the same channel simultaneously.
The network is time-slotted with global synchronization.
At each round $t$, node $v_i$ has $M$ choices of channels, where
channel $c_j$ having data rate drawn from an i.i.d stochastic process $\xi_{i,j}(t)$ over time with a mean $\mu_{i,j} \in[0,1]$. Without loss of generality, we assume the same channel may demonstrate different channel quality for different users. For the same channel $c_j$, the random process $\xi_{i,j}(t)$ is independent from $\xi_{i',j}(t)$ if $i \neq i'$.

At each round $t$, an $N$-dimensional \emph{strategy} vector $\mathbf{s}_x(t)=\mathbf{s}_x=\{s_{x,i}|i=1,\dots,N\}$ is selected under some \emph{policy} from the \emph{feasible strategy set} $F$. Here $s_{x,i}$ is the index of channel selected by node $v_i$ in strategy $\mathbf{s}_x$. We use $x=1,\dots,X$ to index strategies of feasible set $F$ in the decreasing order of average throughput $\lambda_x=\sum_{i=1}^{N} \mu_{i,s_{x,i}}$.
By feasible we mean that all nodes can transmit simultaneously without conflict.
When a strategy $\mathbf{s}_x$ is determined, each node $v_i$ observes the data rate ${\xi}_{i,s_{x,i}}(t)$ of its selected channel, and then the total throughput of the network at $t$ is defined as,
 $
    R_x(t)= \sum_{i \in \textbf{s}_x}{\xi}_{i,s_{x,i}}(t).
 $
We evaluate policies using \emph{regret}, which is defined as the
difference between the expected throughput that could be obtained by a
static optimal policy with the existence of a genie, and that obtained
by the given policy. Let $R_1 = \lambda_1$ be the optimum fixed channel
access strategy, then regret can be expressed as
{\small{
\begin{eqnarray}
    \mathfrak{R}(n)= nR_1 - E\biggl[\sum_{t=1}^{n} R_x(t)\biggl]
    = \sum_{x:R_{x} < R_1} \Delta_x E\bigl[T_x(n)\bigl]
\end{eqnarray}
}}

%% file: formulation.tex
\begin{table}[t]\setlength{\tabcolsep}{3pt}
\begin{center}
\caption{ Summary of notations}
\label{notations}
\begin{tabular}{c| c}
\hline
Variable & Meaning \\
  \hline
    $J_{G,r}(v)$     & $r$-hop neighborhood of node $v$ in $G$         \\
    $J_{H,r}(v)$     & $r$-hop neighborhood of vertex $v$ in $H$         \\
    $A_{r}(v)$        & set of Candidate vertexes in $r$-hop neighborhood of $v$   \\
    $\textrm{MWIS}(I)$ & maximum weighted independent set for vertex set $I$  \\
    $\textrm{MIS}(I)$ & independent set with maximum cardinality for vertex set $I$ \\
    $W(I)$           & summed weight of all vertexes in vertex set $I$\\
    $\mathbf{s}_{x}(t)$ & strategy decision for round $t$ \\
  \hline
  \end{tabular}
\end{center}
\end{table}

\begin{figure}[t]
\centering
    \includegraphics[width=6cm]{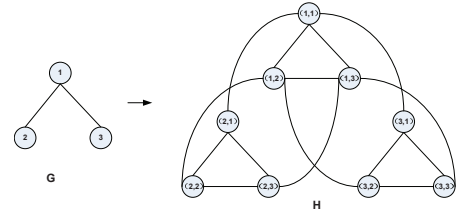}
\caption{Original conflict graph G to extended conflict graph H}
\label{gtoh}
\vspace{-20pt}
\end{figure}

We first analyze the optimum throughput on the assumption that the  mean of each random variable is known.
 We remodel the network $G=(V,E,C)$ as an extended conflict graph
 $H=(\widetilde{V},\widetilde{E})$, where $\widetilde{V}=\{v_{i,j}
 \mid i\in[1,N],j\in[1,M]\}$, and show that the problem can be reformulated as a MWIS problem in extended conflict graph $H$.
Define a set of virtual vertices $\{v_{i,j}$, $j=1,\dots,M\}$ for each
node $i$ and connect $v_{i,j}$ with $v_{i,k}(j\neq k)$ for all
$j,k$. Node $v_i$ is master node of virtual vertex $v_{i,j}$, while
$v_{i,j}$ is slave of $v_i$.
Connect $v_{i,j}$ with $v_{p,j}$  if $i$ and $p$ has an edge in
original network $G$. Then  graph $H$ has $N\times M$ vertexes. We
give an instance in Fig. \ref{gtoh} where the original network
$G$ has $3$ available channels and $3$ nodes.

As each node of $G$ has a clique of virtual vertexes in $H$, and
vertexes with the same channel index  retain the conflict
relationships of master nodes in $G$, then it is straightforward that
a MWIS of $H$ is a throughput-optimal allocation of channels in
$G$. Indeed, an IS of $H$ one-to-one maps to a feasible strategy in
$F$.
Therefore, the feasible strategy set $F$ consists of all independent sets (\textbf{IS}) of vertexes in $H$.
Here note that the independence number of $H$ is less than $N$ if the
chromatic number of $G$ is greater than $M$, and is $N$ otherwise.
The actual length of a feasible strategy may be smaller than $N$ 
if some nodes do not choose any channel.
Let $\xi_{i,j}(t)$ be weight of virtual vertex $v_{i,j}$. 
If the mean of $\xi_{i,j}(t)$ is known, the optimum strategy is to
 find a maximum weighted independent set of vertexes from $H$ as choices made by nodes in $G$, i.e,
{\small{
\begin{eqnarray}
   R_1 = \max_{\mathbf{s}_x\in F} \sum_{i=1}^{N} \mu_{i,s_{x,i}} \nonumber\\
      \mbox{s.t. } F \mbox{ is feasible strategy set.}
\end{eqnarray}
}}
However, these random variables  are unknown actually, each user needs to learn and estimate the weight of each strategy, denoted by $W_x(t)=\sum_{x_i \in \mathbf{s}_x} w_{i,s_{x,i}}(t)$, where $w_{i,s_{x,i}}(t)$ is estimated weight of random variable $\xi_{i,s_{x,i}}(t)$.
Thus, our problem   becomes a NP-hard combinatorial multi-armed bandits problem that selects at most $N$ arms (i.e., vertexes in $H$) out of $K=N M$ ones to minimize the regret $\mathfrak{R}$, such that these arms are independent from each other in $H$.
For brevity, we map the channel index $s_{x,i}$ of node $v_i$ to arm index  $k=(i-1)N+s_{x,i}$.

For NP-hard combinatorial multi-armed bandits problems, a weaker
vision of regret, called \emph{$\beta$-regret} \cite{my2013mab},  is
defined as the difference between the expected throughput that is
$1/\beta$ of the optimum, and that gained throughput (a
$\beta$-approximation policy which instead yields a strategy with
learned weight at least $1/\beta$ of the maximum possible weight) .
Let $R_{\beta,x}(t)$ be the reward of strategy $\mathbf{s}_{\beta,x}$  generated by  the $\beta$-approximation policy, then $\beta$-regret can be expressed as
{\small{
\begin{eqnarray}
    \mathfrak {R}_{\beta}(n)
    = \sum_{R_{\beta,x} < R_1/{\beta}} \Delta_{\beta,x} E\bigl[T_{\beta,x}(n)\bigl]
    +  \sum_{R_{\beta,x} \ge R_1/{\beta}} \Delta_{\beta,x} E\bigl[T_{\beta,x}(n)\bigl] \nonumber
\end{eqnarray}
}}
where $T_{\beta,x}(n)$ is the number of times that strategy $\mathbf{s}_{\beta,x}$ has been played by round $n$, and $\Delta_{\beta, x}$ is the  distance between $R_1/{\beta}$ and mean throughput of strategy $\mathbf{s}_x$.

Though a previous learning policy in \cite{gai2012mab} achieves
zero-regret, the upper bound of regret heavily depends on the
distribution of strategies in feasible set $F$ (or $F_{\beta}$ which
is all strategies whose throughput is at least $R_1/\beta$). That is,
the upper bound of regret $\mathfrak{R}$ (or $\beta$-regret $\mathfrak
{R_{\beta}}$) including a factor of $\frac{1}{\Delta_{\min}}$ (or
$\frac{1}{\Delta_{\beta,\min}}$) becomes vacuous if $\Delta_{\min}$ (or
$\Delta_{\beta,\min}$) $\rightarrow 0$.
We expect  a zero-regret policy without dependency on $\Delta_{\min}$.
Meanwhile,  it admits distributed implementation with low computation
and communication complexity to guarantee efficiency of the channel
access process.

%% file: channelaccess.tex
Each round is divided into two sequent parts, one for strategy decision and the other for data transmission.
In the strategy decision part, it utilizes the learned information in history to determine which strategy shall be selected for current time.
In the data transmission part, users access  corresponding channels to transmit data, and observe real data rate after transmission.
We assume   a common control channel for control message passing in strategy decision.

%% file: channelaccess-sub-learning.tex
To learn for the best possible strategy, we adopt the learning policy proposed in \cite{my2013mab}
where the upper bound of regret is independent with $\Delta_{\min}$ (or
$\Delta_{\beta,\min}$).
The centralized form of the learning policy is shown in Algorithm~\ref{alg1},
where in (\ref{e-max}) it utilizes estimated weight of each vertex in
$\widetilde{V}$  to select a maximum weighted independent set as
strategy decision for next channel access. The estimated value for
actual weight $\xi_{i,j}(t+1)$ of vertex $v_{i,j}$ is
{\small{
\begin{equation}
\label{weight}
    w_{i,j}(t+1)= \tilde{\mu}_{i,j}(t)  +\sqrt{\frac{\max{(\ln{\frac{t^{2/3}}{K m_{s_{x,i}}} }},0)} {m_{s_{x,i}}}}.
\end{equation}
}}

\begin{algorithm}[tbhp]
\caption{Learning policy}
\label{alg1}
  \begin{algorithmic}[1]
   \STATE For each round $t$,
    select a strategy $\mathbf{s}_x$ by maximizing
        {\small{
        \begin{equation}
        \label{e-max}
          \max_{\mathbf{s}_x\in F} \sum_{s_{x,i} \in \mathbf{s}_x}
             \biggl( \tilde{\mu}_{s_{x,i}}(t)
                    +\sqrt{
                         \frac{ \max{(\ln{\frac{t^{2/3}}{K m_{s_{x,i}}} }},0)} {m_{s_{x,i}}}
                           }
             \biggl).
        \end{equation}
        }}
  \end{algorithmic}
\end{algorithm}

Consequently, it only requires to store and update estimation for $MN$ vertexes that  costs storage and computation linear with $MN$, instead for $M^N$ strategies in $F$ that costs storage and computation linear with $M^N$.
More specially, we need two $1\times K$ vectors to store and update the estimated weight. One is $(\tilde{\mu}_k)_{1\times K}$ where
$\tilde{\mu}_{k}$ is  observed mean of $\xi_{k}$ up to the current round, and the other is $(m_{k})_{1\times K}$ where $m_k$ is the number of times that channel $\xi_{k}$ has been selected so far.
After data transmission on the channels of chosen strategy
$\mathbf{s}_{x}$ in slot $t$,   actual weight $\xi_{s_{x,i}}(t)$ is
observed for all $s_{x,i} \in \mathbf{s}_{x}$. Then
$(\tilde{\mu}_k)_{1\times K}$ and $(m_{k})_{1\times K}$ are updated in
the following way:
{\small{
\begin{eqnarray}
\label{update-mu}
&\tilde{\mu}_k(t) = \left\{ \begin{array}{ll}
\frac{\tilde{\mu}_k(t-1) \cdot m_k(t-1)+\xi_k(t)}{m_k(t)}& \textrm{if $k\in \mathbf{s}_x$}, \\
\tilde{\mu}_k(t-1) & \textrm{else.}
\end{array} \right. \\
\label{update-m}
&m_k(t) = \left\{ \begin{array}{ll}
m_k(t-1)+1 & \textrm{if $k\in \mathbf{s}_x$},\\
m_k(t-1) & \textrm{else}.
\end{array} \right.
\end{eqnarray}
}}

 Due to NP-hardness of the MWIS problem in (\ref{e-max}), it is desirable to solve it   approximately while retaining zero-regret.
The following theorem  shows that, for
\emph{any} algorithm with
approximation ratio at least  $1/\beta$ for the MWIS problem,
the regret on the achieved throughput is bounded.

\begin{theorem}\cite{my2013mab}
    The $\beta$-approximation learning policy has
    {\small{
    \setlength\arraycolsep{2pt}
    \begin{eqnarray}
     \sup \mathfrak{R}_{\beta}(n)
                    &\leq&  \frac{1}{\beta}  N K + \left( \sqrt{e K} +  \frac{16}{e \beta}( 1+N)N^3 \right) n^{\frac{2}{3}} \nonumber \\
    \label{betaregret}             &+&     \frac{1}{\beta} \left( 1+ \frac{4 \sqrt{K} N^2}{e \beta^2}  \right)N^2 K n^{\frac{5}{6}}
    \end{eqnarray}
    }}
      without dependency on $\Delta_{\beta,\min}$. The supremum is taken over all $X$-tuple of probability distributions on $[0,1]$.
\end{theorem}

%% file: channelaccess-sub-centralized.tex
Intuitively, the greater the  value $\beta$ is, the more sacrifice on
overall throughput it causes. Given that, we choose the robust PTAS
proposed in \cite{nieberg2005robust} to solve the MWIS problem. Though
centralized, the robust PTAS is elegant, and more importantly, it
requires no geometric information as other PTAS schemes
\cite{mwis2005} \cite{intersection2010}. This feature is very
attractive as it is expensive to get and maintain exact locations of
each node in multi-hop wireless networks, not to say negative effect
of  errors  by location methods.
We will show how to implement it in a distributed manner later. For
better understanding, we first introduce the centralized method.

\textbf{Robust PTAS}.
We begin with some notations.
Given a unit disk graph $G=(V,E)$ with a set ${V}$ of nodes  and a set $E$ of edges,
an edge $(u,v) \in E$ if the Euclidean distance $\|u,v\|\leq 2$.
For a subset $I$ of nodes in $V$, let $W(I)$ denote the total weight
of $I$, i.e., $W(I)=\sum_{v_i\in I}w_i$, and $\textrm{MWIS}(I)$ denote
a maximum weighted independent set for $I$. The independent set with
maximal cardinality (\textbf{MIS}) for $I$ is written as
$\textrm{MIS}(I)$.
Let $d_G(u,v)$ be the minimum hop of any path connecting $u$ and $v$
in $G$.
Define
\begin{equation*}
       J_{G,r}(v):=\{  u \in V \mid  d_{G}(u,v) \le r \}
\end{equation*}
be the \emph{$r$-hop neighborhood} of $v$ in $G$.
The \emph{$r$-hop distance of $G$}, $L_{G,r}(v)$ is the maximum
Euclidean distance between $v$ and neighbors in $J_{G,r}$.
Clearly $L_{G,r}(v) < 2r$.

Let $\epsilon > 0$ and $\rho := 1+\epsilon$ denote the desired
approximation guarantee.
In graph $G$, the algorithm starts with a node of maximal weight
$w_{\max}=\{\max w_v| v \in {V}\}$, and then computes
$\textrm{MWIS}(J_{G,r})$ as long as $W(\textrm{MWIS}(J_{G,r+1})) >
\rho W(\textrm{MWIS}(J_{G,r}))$  holds. Let $\bar{r}$ denote the
smallest $r$ for which the criterion is violated.
It has been proved that $\bar{r}$ is a constant for a
specific $\rho$, i.e., $\rho^r\leq(2r+1)^2$.
We then remove $\textrm{MWIS}(J_{G,\bar{r}}(v_{\max}))$ and all the adjacent
vertices from $G$, and repeat the above process on the remaining graph.
Then the union of all removed independent sets form an
 independent set, and it is proved that it is $\rho$-approximation for
 the MWIS of unit disk graph $G$.

As the extended conflict graph $H$ is not a strict unit disk graph, we
distinguish some notations.
Define \emph{$r$-hop neighborhood in extended graph $H$} as
\begin{equation*}
        J_{H,r}=J_{H,r}(v):=\{  u \in \widetilde{V}    \mid  d_{H}(u,v)  \le r \}.
\end{equation*}
Note that two vertexes that belong to the same master node of $G$ has $0$ Euclidean distance geometrically, but they are $1$-hop neighbors in $H$.
The \emph{$r$-hop distance of $H$}  also satisfies  $L_{H,r}(v)=L_{H,r} < 2r$.
We then have the following theorem on approximation ratio achieved by robust PTAS in $H$.

\begin{theorem}
    \label{th:robustPTASinH}
    Robust PTAS applies to extended conflict graph $H$ with
    approximation ratio $\rho$, where $\rho^r=M\cdot (2r+1)^2$.
\end{theorem}
\begin{IEEEproof}
 Robust PTAS can be equally extended to other intersection graphs as long as the graph is growth-bounded,
  where the number of independent vertexes in a vertex's $r$-hop neighborhood is constantly bounded\cite{kuhn2005fast}\cite{nieberg2005robust}. Though the extended graph $H$ is not a strict unit graph, it is straightforward to verify that $H$ is growth-bounded.
Note that a set of virtual vertexes that belong to the same master
node form a clique in $H$.
For a node $v_i \in V$ in $G$, the independent number of
$J_{G,r}(v_i)$ is upper bounded by $(2r+1)^2$.
As each vertex in $G$ will define $M$ slave vertexes in $H$, a simple
pigeonhole principle shows that the number of independent vertexes in
the $r$-hop neighborhood $J_{H,r}(v)$ of graph $H$ is bounded from
above by $M\cdot (2r+1)^2$.
 Thus we say $H$ is also growth-bounded, and the approximation ratio
 achieved in $H$ satisfies $\rho^r \leq M \cdot (2r+1)^2$.
\end{IEEEproof}

%% file: channelaccess-sub-distributed.tex

\begin{algorithm}[tb]
\caption{Main framework of distributed channel access}
\label{framework}
{\small
  \begin{algorithmic}[1]
    \FOR{Round $t=1,\dots,n$, $\forall v \in \widetilde{V}$}
        \IF {$v$ belongs to strategy decision $\mathbf{s}_x(t-1)$ in previous round}
            \STATE Broadcast its new weight in $(2r+1)$-hop neighborhood.
        \ENDIF
        \STATE Receive  all updated weight on $(2r+1)$-hop neighbors,
                and update corresponding weight.
        \STATE Perform  distributed Robust PTAS as Algorithm~\ref{distributed}
                within $D$ mini-rounds.
        \IF{$v$ is marked as \emph{Winner} }
                \STATE Access the channel to transmit data.
                \STATE Observe actual data rate.
                \STATE Update estimated weight using Equation (\ref{update-mu}), (\ref{update-m}) and (\ref{weight}).
        \ENDIF
    \ENDFOR
  \end{algorithmic}
}
\end{algorithm}


\begin{figure}[t]
\centering
    \includegraphics[width=8.5cm]{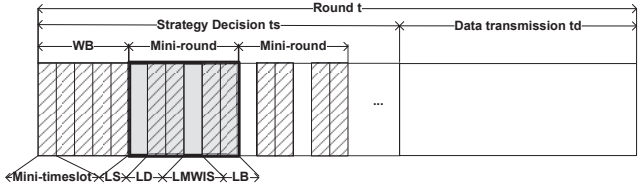}
        \caption{Structure of a single round:\textbf{WB}-weight update;
                \textbf{LS}-LocalLeader selection;
                \textbf{LMWIS}-local computation of MWIS;
                \textbf{LB}-local broadcast of status determination.}
        \label{round}
\vspace{-0.15in}
\end{figure}



    As the centralized form of robust PTAS algorithm requires centralized
computation and global collection of weight/observed information, it
costs high computation (i.e., $O(N^{\rho^r})$) and communication
complexity that is unwelcome in multihop networks. We design a
distributed implementation that takes low communication and
computation complexity.

The main framework of our distributed implementation is shown in Algorithm~\ref{framework},
    which is run round by round,
    where each \emph{round} consists of a strategy decision part
    and a data transmission part (see Fig.~\ref{round}).
The strategy decision part includes an initiation step called Weight Broadcast (\textbf{WB}),
    where each vertex broadcasts its new weight if it accessed channel in previous round
    (i.e., included in previous strategy decision $\mathbf{s}_x(t-1)$),
    to ensue computation of MWIS with newest weight.
In our protocol, these vertexes in   $\mathbf{s}_x(t-1)$
    broadcast updated weight information within hops $(2r+1)$ to ensure independence of the final output, for which
    we will explain later.
Let \emph{mini-timeslot} be the time unit required
    for a round of communication between two connected vertexes.
    In the first round, the initial weight of each vertex is $0$,
so  vertexes can be randomly selected as LocalLeader,
or they can use their IDs as weight.
    In the later case, it will cost $O(N)$ mini-timeslots
to  collect IDs of all neighbors even in a local neighborhood.
    In next rounds, however, it costs only $O((2r+1)^2)$
mini-timeslots to finish the WB process.
    The key observation is that within any $(2r+1)$-hop neighborhood of any vertex,
at most $O((2r+1)^2)$ vertexes are selected as independent vertexes.
    Only independent vertexes selected in a strategy decision observe  new values,
and utilize the observation to update estimated weight
(i.e., plugging~(\ref{update-mu}) and (\ref{update-m}) into~(\ref{weight})).
    If each vertex performs weight broadcast sequently,
obliviously it will take  $O((2r+1)^3)$ mini-timeslots
to finish the whole procedure in a $(2r+1)$-hop neighborhood.
    As an alternative, these selected vertexes can
efficiently broadcast their weight using pipeline methods
such as constructing a connected dominating set
\cite{huang2008broadcast} \cite{wang2005cds} \cite{zou2011cds},
by which number of mini-timeslots can be reduced to $O((2r+1)^2)$.

    After WB,  each vertex  then runs distributed Robust PTAS presented Algorithm~\ref{distributed}
to compute MWIS with updated weight. In our protocol, we will run $D$ \emph{mini-rounds}
to output a final IS with a good approximation ratio to the optimum.
    When finishing execution of Algorithm~\ref{distributed}, the vertexes included
in current strategy decision access channels for data transmission, where they obtain
new observation  to update estimation of weight  for the next round.
    Until now a full round of Algorithm~\ref{framework} completes, and a new round follows.



\begin{algorithm}[tb]
\caption{Distributed robust PTAS for strategy decision at each vertex}
\label{distributed}
{\small
  \begin{algorithmic}[1]
    \REQUIRE {$\forall \textrm{ vertex } v \in \widetilde{V}$, marked as status \emph{Candidate},
                have collected newest weights of all $(2r+1)$-hop neighbors ${J}_{H,2r+1}(v)$}.
    \FOR{mini-round $\tau=1,2,\dots, N$ }
        \IF {$v$ is \emph{Candidate}}
            \IF {$w_v \geq \max\{w_u | u \in {A}_{2r+1}(v) \}$}
                \STATE $v$ is marked as \emph{LocalLeader} and declare in $(2r+1)$-hop neighborhood.
            \ENDIF
        \ENDIF
        \IF{$v$ is \emph{LocalLeader}}
            \STATE Compute a local $\textrm{MWIS}( A_{r}(v) )$   using
            enumeration.
            \STATE Determine status of $r$-hop neighbors. For any
            \emph{Candidate} vertex in ${A}_{r}(v)$, marked as
            \emph{Winner} if it is in $\textrm{MWIS}( A_{r}(v)) $, or
            marked as \emph{Loser} otherwise.
            \STATE Locally broadcast the results  within $(3r+1)$-hop
            neighborhood of the LocalLeader.
            \STATE Update its own status accordingly.
        \ELSIF{$v$ is \emph{Candidate}}
            \IF{$v$ receives determination messages}
                \STATE Update status of itself and $(2r+1)$-hop
                neighbors accordingly.
            \ENDIF
          \ENDIF
    \ENDFOR
     \end{algorithmic}
}
\end{algorithm}

\begin{figure}
\centering
    \includegraphics[width=8cm]{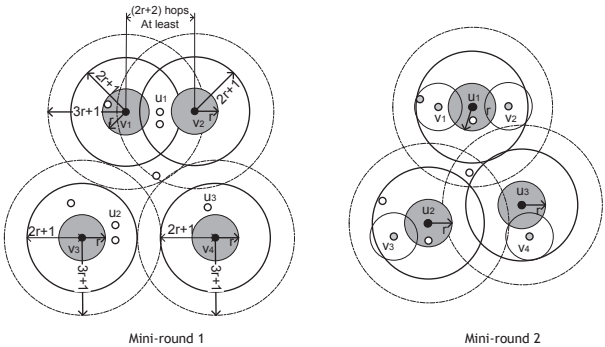}
        \caption{Illustration of Algorithm \ref{distributed}
                    in two sequent mini-rounds, where vertexes $v_i,i=1,2,3,4$
                    are selected as LocalLeader at mini-round 1, and vertexes $u_i,i=1,2,3$
                    become LocalLeader at mini-round 2 after neighbors with bigger weight excluded. }
        \label{miniround}
\vspace{-0.2in}
\end{figure}


 Now we describe distributed Robust PTAS in Algorithm~\ref{distributed}.
 We introduce four statuses in Algorithm~\ref{distributed}: \emph{Candidate},
 \emph{LocalLeader}, \emph{Winner} and \emph{Loser}.
A \emph{Candidate} is one vertex that is not marked as Winner or
Loser, and thus has opportunity to be a Winner.
 Initially, at the start of each round, each node is marked as
 \emph{Candidate}.
A \emph{LocalLeader} is a Candidate that has the maximum weight among
 all its Candidate neighbors in $(2r+1)$-hop neighborhood.
Each LocalLeader will compute the maximum weighted independent set
 using all \emph{Candidate} vertexes in its $r$-hop neighborhood.
A \emph{Winner} is a vertex that is included in the final resulting IS
 computed from LocalLeader,
 while a \emph{Loser} is a vertex that is neither Candidate nor
 Winner.
Notice that here we use the $(2r+1)$-hop neighborhood to find a
 \emph{LocalLeader} while use $r$-hop neighborhood to compute an IS.
This approach will assure that the union of all the independent sets
 computed by all selected LocalLeaders form an independent set,
 as the hop-distance between any two LocalLeaders is at least $2r+2$
 and the hop distance between any two vertexes from the computed
 independent sets by two LocalLeaders is at least $2$.

\begin{figure}[t]
\centering
    \includegraphics[width=8cm]{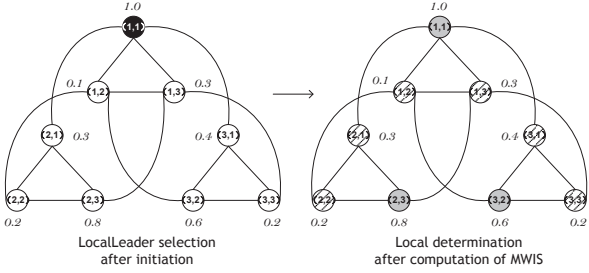}
\caption{Local computation of Algorithm \ref{distributed} in extended graph $H$ with estimated weight: black vertex-\emph{LocalLeader}, white vertex-\emph{Candidate}, stripe vertex-\emph{Loser}, gray vertex-\emph{Winner}. }
\label{algorithm_2}
\vspace{-20pt}
\end{figure}

    Let ${A}_{r}={A}_{r}(v)$  be the set of all Candidate vertexes in  $J_{H,r}(v)$
to exclude vertexes that have been marked as Winner or Loser.
    The algorithm begins with the process called LocalLeader selection (Line $2-6$).
To ensure independency of the union of all local computed results,
each LocalLeader compute local MWIS within $r$-hop neighborhood.
    A LocalLeader has to broadcast its computed MWIS results  among
$(3r+1)$-hop neighborhood (Line $10$), so that Candidate vertexes in
the next round have complete status information on its $(2r+1)$-hop
neighbors to correctly continue the algorithm.
    Notice that a Candidate vertex, say $u$, in the current round could become a
LocalLeader in the next round.
    For this to happen, it must be the case that 1) at current round,
there is a virtual vertex, say $x$, within its $(2r+1)$-hop whose
weight is larger, 2) after this round, the virtual vertexes with
larger weight change their status (either they are LocalLeaders or
they are decided by other LocalLeaders as Winner or Loser).
    Thus, to assure correct operation, the status of a virtual vertex, say
 $u$, should be broadcast by its LocalLeader, say $v$, to the $3r+1$
hops, as the hop distance between $u$ and $v$ could be as large as $3r+1$.
    For better understanding, we illustrate distributed execution
of Algorithm~\ref{distributed} in two sequent mini-rounds in Fig.~\ref{miniround},
and local computation in a single mini-round in Fig. \ref{algorithm_2}
for the network presented in Fig. \ref{gtoh}.

    At each mini-round, vertexes either marked as Winner or Loser will be excluded
and stop executing the algorithm. The algorithm terminates when no candidates exist,
i.e., all vertexes are marked as either Winner  or Loser,
which may require mini-rounds $O(N)$.
    Actually, a constant number of mini-rounds is enough to output a good decision.
    That is why we set $D$ mini-rounds of Algorithm~\ref{distributed} in
main framework Algorithm \ref{framework}.

    Herein we first  present the achieved approximation ratio
after $O(N)$ mini-rounds. The results on mini-rounds $D$ will
be presented in later analysis.
\begin{theorem}
    Algorithm \ref{distributed} achieves the same approximation ratio $\rho$ as the centralized robust PTAS in $H$.
\end{theorem}
\begin{IEEEproof}
    Let $v_i(\tau)$ be a LocalLeader selected at mini-round $\tau$.
    In each mini-round, a LocalLeader $v_i(\tau)$ utilizes the robust PTAS to
find $\textrm{MWIS}(A_{r})$ in its effective $r$-hop
neighborhood.  Thus, each $\textrm{MWIS}(A_{r})$ computed by a
LocalLeader is $\rho$-approximation to $\textrm{MWIS}(A_{r+1})$.
    Let $\textrm{MWIS}(\widetilde{V})$ be the global optimum, and $I$ be
intersection of $\textrm{MWIS}(\widetilde{V})$ and
$A_{r+1}(v_i(\tau))$, we have $ W(I) \leq \rho
W({\textrm{MWIS}(A_{r+1})})$. As union of $A_{r}$ in all mini-rounds is
exactly $\widetilde{V}$ and any two distinct  $A_{r}$ do not
intersect, we have the union of all  $\textrm{MWIS}(A_{r})$ output
by all LocalLeaders is $\rho$-approximation to the global optimum
$\textrm{MWIS}(\widetilde{V}) $ in weight.
\end{IEEEproof}

{\textbf{Complexity}.}
    We summarize complexity in a complete round. \\*
\emph{Communication complexity}: As shown in Fig.~\ref{round},
    local broadcast happens $3$ times in each round,
    respectively for WB, LD, and LB.
WB could be finished within mini-timeslots $O((2r+1)^2)$,
    which costs each vertex  $O((2r+1)^2)$ number of messages in worst case.
LD is done by a LocalLeader in its $(2r+1)$-hop neighborhood, then it costs $O(2r+1)$ mini-timeslots,
    and each vertex  $O(1)$ passing messages.
In LB, each LocalLeader has to broadcast the results within its $(3r+1)$-hop neighborhood.
    There are at most $\frac{2\pi \cdot 2r}{2r+1}=O(1)$ number of LocalLeaders 
     within any $(2r+1)$-hop neighborhood of any vertex.
    Thus it costs mini-timeslots $O(3r+1)$, and communication complexity $O(1)$.
    Totally, it requires mini-timeslots $O(r^2+Dr)$, and each vertex number of passing messages $O(r^2+D)$. \\*
\emph{Computation complexity}:
    The main computation cost is caused by LMWIS, as LS can be finished instantly.
    In every mini-round, we use complete enumeration to compute local  MWIS in each $A_{r}(v)$.
    Suppose there are $m$ nodes in corresponding $r$-hop neighborhood of $G$, then $|A_{r}(v)| \leq Mm$.
    Since $|\textrm{MWIS}(A_{r}(v))| \leq M(2r+1)^2$, there are totally  $C_{Mm}^{M(2r+1)^2} $ enumerations.
    Using $M(2r+1)^2 < \rho^r\textrm{ for } r< \bar{r}$, we have
    {\small
    \setlength\arraycolsep{0.5pt}
    \begin{eqnarray}
        C_{Mm}^{M(2r+1)^2}   \leq  \left( \frac{Mme}{M(2r+1)^2}\right)^{M(2r+1)^2}
        \label{complexity}   \leq  \left( \frac{me}{(2r+1)^2}\right)^{\rho^r}.
    \end{eqnarray}}
    Hence, it requires polynomial time $O(m^{\rho^r})$ per mini-round,  
    and $O(Dm^{\rho^r})$ per round.
    In practice, we can use more efficient constant approximation algorithm instead,
    the communication complexity reduces to $O(D)$ with a worse  approximation ratio.\\*
\indent\emph{Space complexity:}
    It is $O(m)$, as each vertex has to store weight of neighbors within $(2r+1)$-hop neighborhood.

    In our protocol, $r$ and $D$ is constant, then the communication, computation, 
    and space complexity is $O(1)$, $O(m^{\rho^r})$, and $O(m)$ respectively.

\begin{figure}
\centering
    \includegraphics[width=5cm]{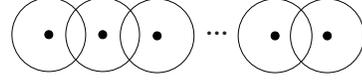}
\caption{Worst case of strategy decision caused by linear effect where $w_1 > w_2 \dots > w_N$}
\label{linear}
\vspace{-20pt}
\end{figure}

%% file: channelaccess-sub-improve.tex
As mentioned previously, the distributed implementation of strategy decision requires $O(N)$ mini-rounds to get all vertexes marked. We then show a simple instance of the worst case  in Fig. \ref{linear}.
In the figure we use a linear network where all vertexes are aligned uniformly along a line within  $1$-hop distance. One can easily figure out that when the weight of each vertex is in a decreasing order from the start vertex to the end vertex, at the beginning only the start vertex can be LocalLeader since no other vertexes are locally maximum weighted. 
And, in each next round, still only one could be LocalLeader sequently.
Thus it would take $N$ mini-rounds in a single round.

We then analyze the time complexity under random networks where location of each vertex is uniformly random distributed.
We assume a random network has an average degree of $d$.
We expect to show it is possible to achieve a slightly smaller constant-approximation ratio if   the algorithm terminates after a fixed number of $D \ll N$ mini-rounds, no matter whether there remains vertexes unmarked or not. Surprisingly, we find that it is indeed the case. The following theorem presents our results.

\begin{theorem}
\label{constant}
     Given a random network $G$ with an average degree $d$, Algorithm \ref{distributed} achieves $\alpha$-approximation to the optimum if we set the number of mini-rounds as a constant $D \ll N$.  $\alpha$ is a constant with constant probability.
\end{theorem}
\begin{IEEEproof}
The proof is omitted here due to space limit.
\end{IEEEproof}
By Theorem~\ref{constant} and (\ref{complexity}), time complexity of Algorithm~\ref{distributed} reduces to  $O(d^{\rho^r})$ regarding to random networks.

%% file: channelaccess-sub-regret.tex
Now we  analyze   \emph{practical regret} (or \emph{effective throughput}) that considers the missed throughput due to time spent on learning.
Let $t_{a}$ and $t_{m}$ respectively be length of a single round and mini-round.
Time for strategy decision and data transmission denoted by $t_s$ and $t_d$. In the strategy decision, supposing it requires $c$ mini-rounds, one for weight update, others for strategy decision,
then $t_{a} = t_s + t_d = c t_m + t_d$.
The actual data rate gained at each round is  $R_{x}(t) \cdot t_d / t_a = \theta R_{x}(t)  $, where $\theta= t_d / t_a$. The actual distance between $R_1/\alpha$ and a strategy $\textbf{s}_x$ is  $R_1/{\alpha} - \theta \lambda_x = \theta \Delta_{\theta\alpha,x} $.
Thus in a round, the more time for learning, the larger regret it will be. In practice, we cannot use very long round as $t_a$ shall be smaller than channel coherence time.

Using $\beta=\theta\alpha$ as the approximation ratio, and $\Delta_{\beta,X} = \theta \Delta_X$ as the maximum distance between the actual mean throughput of $R_1/\theta\alpha$ and $\mathbf{s}_X$,
we obtain the practical regret is less than $\theta \cdot \mathfrak{R}_{\theta\alpha}(n)$ according to \cite{my2013mab},
i.e.,
\begin{theorem}
    The practical regret of Algorithm 2 satisfies
    {\small{
    \setlength\arraycolsep{1pt}
     \begin{eqnarray}
     \sup \theta \mathfrak{R}_{\theta\alpha}(n)
      &\leq&  \frac{1}{\alpha}  N K + \left( \theta \sqrt{e K} +  \frac{16}{e\alpha}( 1+N)N^3 \right) n^{\frac{2}{3}} \nonumber \\
    \label{praticalregret}             &+&     \frac{1}{\alpha} \left( 1+ \frac{4 \sqrt{K} N^2}{e (\theta\alpha)^2}  \right)N^2 K n^{\frac{5}{6}}.
    \end{eqnarray}
    }}
\end{theorem}
Then our channel allocation scheme can guarantee an effective throughput of $R_1/(\theta\alpha)- \theta \mathfrak{R}_{\theta\alpha}(n)$. 

%% file: simulation.tex
\vspace{-0.1in}
Now we conduct simulations for our proposed channel accessing scheme under random networks.
We set three series of simulations to respectively study  efficiency, regret and influence of stale weight.
In all simulations we run Algorithm \ref{distributed} with $r=2$.
We set $8$  types of   channels with data rates (units kbps) 150, 225, 300, 450, 600, 900, 1200, and 1350 respectively \cite{li2012almost}.
Each channel evolves as a distinct i.i.d Gaussian stochastic process over time.
We set each round has length of a unit time slot.
 Referring to a cognitive radio system \cite{li2012almost}, we list the values of time parameters of a round in Table \ref{time}. In strategy decision of each round, we set $t_s = 4 t_m$. Let $t_b$ be time to finish local broadcast and $t_l$ be the total time for local computation (LocalLeader selection and local MWIS computation). We have $t_m = 2t_b+t_l = 250 ms$.
According to Fig. \ref{round}, the actual throughput gained at each round is $\frac{t_d}{t_d+4t_m} R_x(t)=0.5 R_x(t)$ in our setting.

\begin{table}[!tp]\setlength{\tabcolsep}{4pt}
\begin{center}
\caption{PARAMETER VALUES FOR SIMULATION}
\label{time}
\begin{tabular}{c |c| c| c}
  \hline
  \hline
   round  $t_a$   & $2000$ms       & local broadcast $t_b$ & $100$ms       \\
   \hline
  local computation $t_l$ & $50$ms    &  data transmission $t_d$ &$1000$ms \\
  \hline
  \end{tabular}
\end{center}
\vspace{-0.2in}
\end{table}

\subsection{Efficiency of Algorithm \ref{distributed}}
\begin{figure}[t]
{\small{
\centering
    \includegraphics[width=8cm]{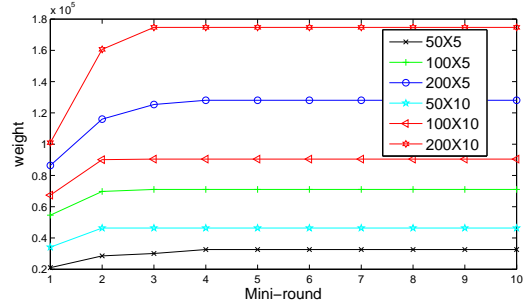}
\caption{Summed weight of all output ISs  as increasing of mini-round under various random networks with varying $N \times M$}
\label{algorithm2test}
}}
\vspace{-30pt}
\end{figure}

  We first set a series of experiments to show efficiency of Algorithm \ref{distributed}.
  We plot the summed weight of all output MWISs by mini-round $1$ to $N$ for various $N \times M$ random networks.
  The value of $N \times M$ is respectively set as
$50 \times 5$, $100 \times 5$, $200 \times 5$, $50 \times 10$, $100 \times 10$, and $200 \times 10$.
  From the Fig. \ref{algorithm2test}, we can see that every line converges to a fixed value after the $4$th mini-round,
no matter how many vertexes there are in the extended graphs.
  This indicates that all vertexes are marked by that time.
  The results coincide with Theorem \ref{constant} where we claim
that our proposed algorithm converges to a constant approximation ratio
that is almost optimal under random graphs.

\subsection{Regret analysis}

    In the second series of experiments, we study practical regret and  $\beta$-regret
caused by  our proposed distributed learning scheme.
    We compare our method with LLR learning policy \cite{gai2012mab}.
    According to definition of regret and $\beta$-regret,
we need to compute the optimal throughput gained by the static best channel allocation.
    As the MWIS problem is NP-hard to solve,
we construct a small network where we could find the optimum by brute force easily.
Here we randomly generate a connected network with $15$  users, each having $3$ channels available.
Using mean date rate of each channel as weight,
we  obtain the weight of the resulting MWIS or optimal throughput of the network, i.e., $7282.90$.

\begin{figure}[htpb]
    \centering
    \subfigure[Practical regret]{
        \includegraphics[height=1.2in,width=1.6in]{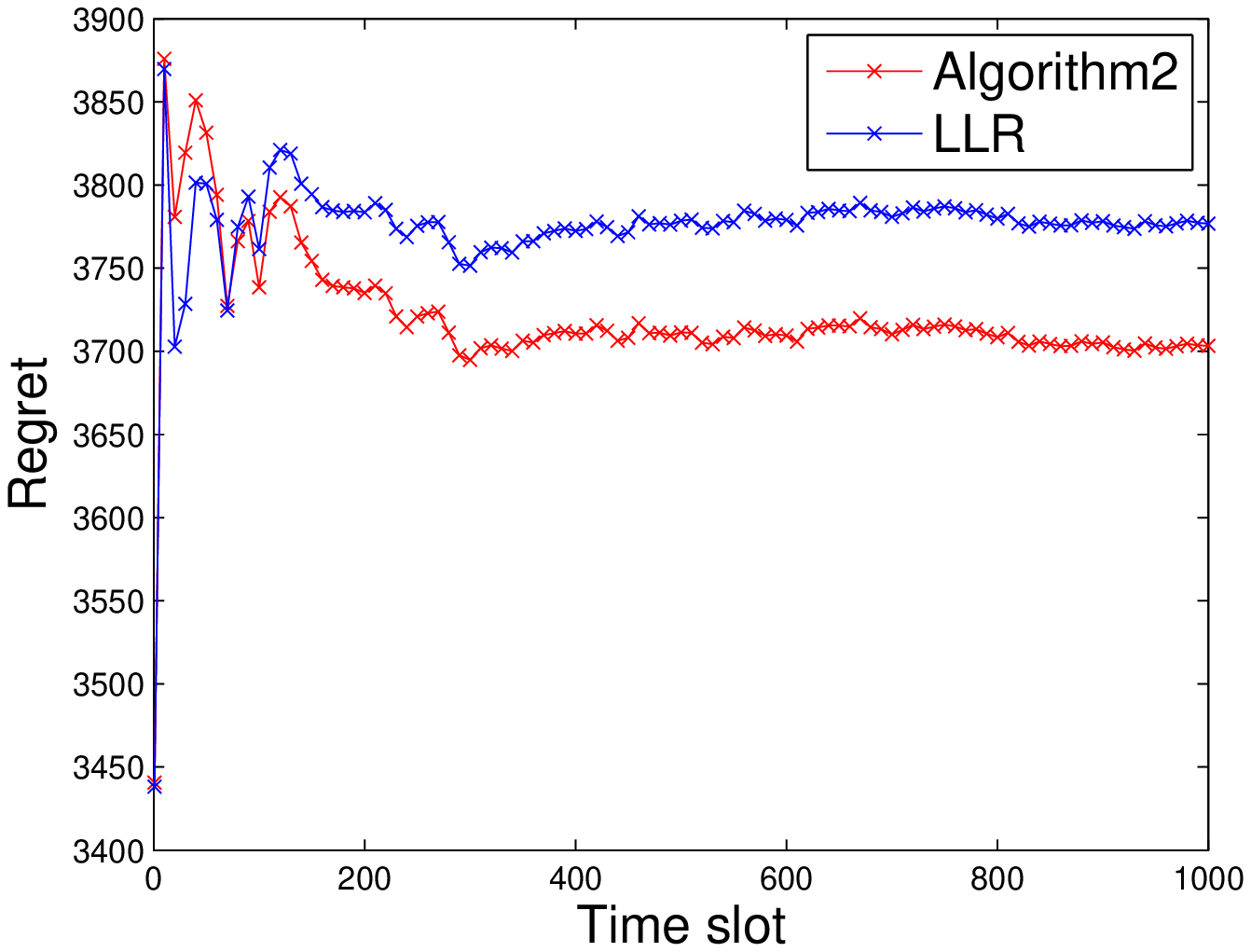}}
  \hspace{0cm}
  \hfill
    \subfigure[Practical $\beta$-regret]{
        \includegraphics[height=1.2in,width=1.6in]{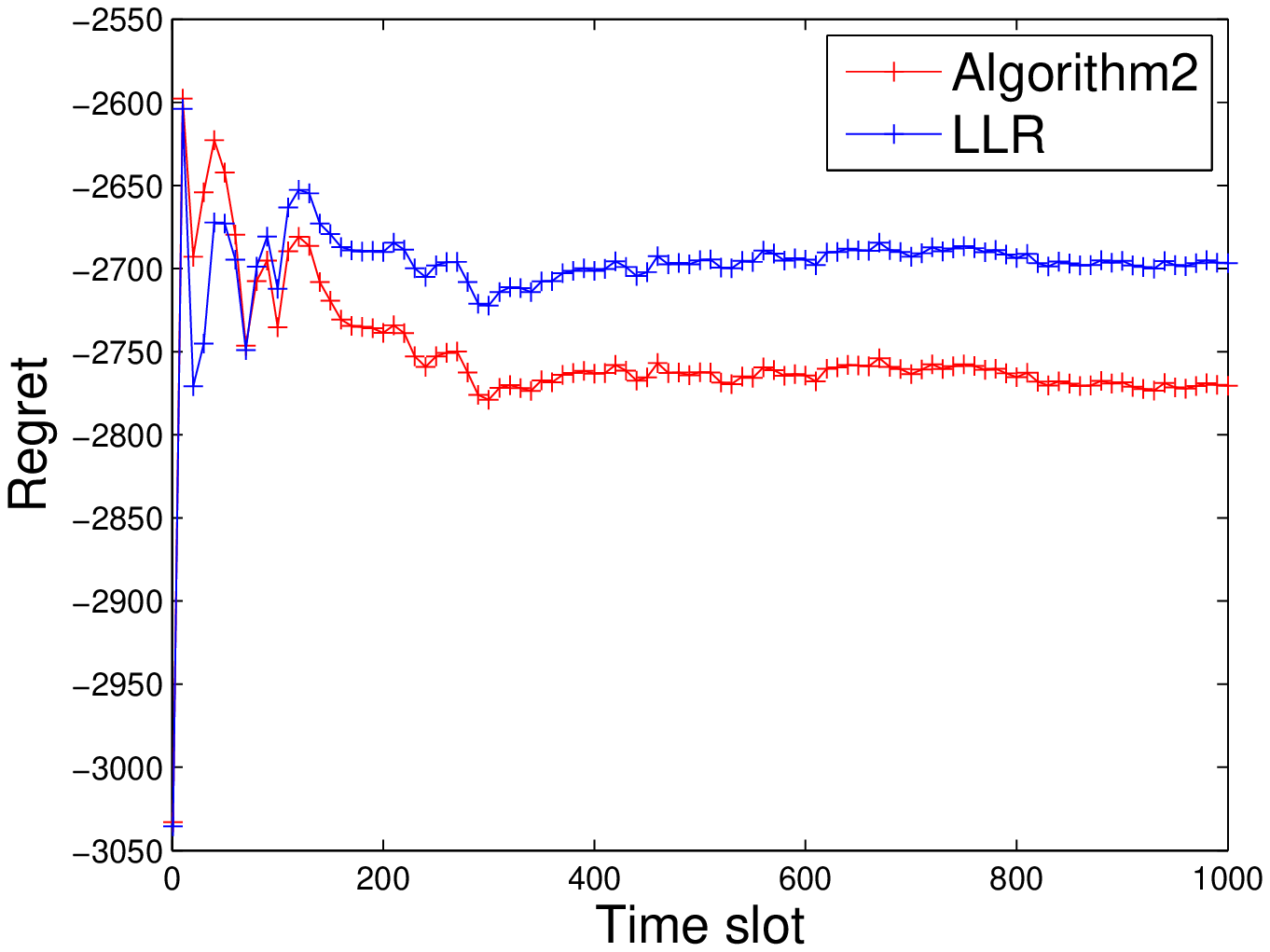}}
   \caption{Practical regret/$\beta$-regret with every-time-slot update: comparison with LLR learning policy  }
 \label{regret}
 \vspace{-10pt}
\end{figure}

    We then compare the optimal throughput, $1/\beta$ of the optimal throughput,
with the effective throughput gained by the two learning algorithms.
    The results  are shown in Fig. \ref{regret},
which plots changes of practical regret and $\beta$-regret as time increases.
    In both figures, our proposed algorithm outperforms the LLR learning policy.
    However, the practical regret compared to the optimum is far beyond $0$,
which indicates a significant impact caused by the time on learning.
    The ideal regret without practical consideration will tend to $0$
as the effective throughput is only half of the observed throughput in our setting.
    As to practical $\beta$-regret, recall that when the reward of selected strategy is greater
than $1/\beta$ of the best reward, the corresponding regret is negative.
    Fig. \ref{regret} (b) also shows that the $\beta$-regret converges to a negative value,
indicating that the achieved throughput by both algorithms is much better than $1/\beta$ of the optimum,
even considering missed throughput on learning.

\subsection{Throughput performance under unfrequent update}
  We evaluate the effective throughput  under different frequencies
of weight update in the third series of simulation,
where meanwhile we  compare performance of our learning policy with LLR policy.
  In our proposed algorithm, initially each vertex has to collect weight
of neighbors inside $(2r+1)$-hop neighborhood.
  If weight as well as corresponding  strategy decision is updated at every time slot, 
  it will cause high communication and communication cost
that significantly affects effective throughput of data transmission.
    Instead, we can update weight every period $P$ that consists of $y$ time slots.
  Then we just need to do strategy decision at the beginning,
and repeat data transmission $y$ times.
  The length of a period is $t_P=y t_a$.
  The actual average throughput gained at the $z^{\textrm{th}}$ period is
{\small{
$
  R_P(z)   = \frac{ R_x(zy+1) t_d  + \sum \limits_{t=zy+2}^{(z+1)y} R_x(t) t_a}  {y t_a}.
$
}}
  We conduct experiments in a random network with $100$ users and $10$ channels.
  For such a large scale network, we will not compute the best static strategy
as it can not be finished instantly.
  Instead, we record the average observed throughput up to $z$ period $ \widetilde{R}_P(z)$,
where
{\small{
$
     \widetilde{R}_P(z) = \frac{ (z-1) \widetilde{R}_P(z-1) + R_P(z)}{z},
$
}}
 and average estimated   throughput  $\widetilde{W}_P(z)$
(i.e., average estimated weight of all selected strategies throughput up to $z$).
  Let  $W_P(z)$ be average estimated  throughput at $z^{\textrm{th}}$ period,
we have
{\small{
$
   W_P(z) = \frac{[(y-1)t_a + t_d] W_x(zy+1)} {y t_a},
$
and
$
    \widetilde{W}_P(z) = \frac{ (z-1) \widetilde{W}_P(z-1) + W_P(z)}{z}.
$
}}
  The difference between $ \widetilde{R}_P(z) $ and $ \widetilde{W}_P(z)$ can
also indicate the throughput performance of the algorithm.

\begin{figure*}[htpb]
    \centering
    \subfigure[1 time slot every period ]{
        \label{p1}
        \includegraphics[height=1.2in,width=1.6in]{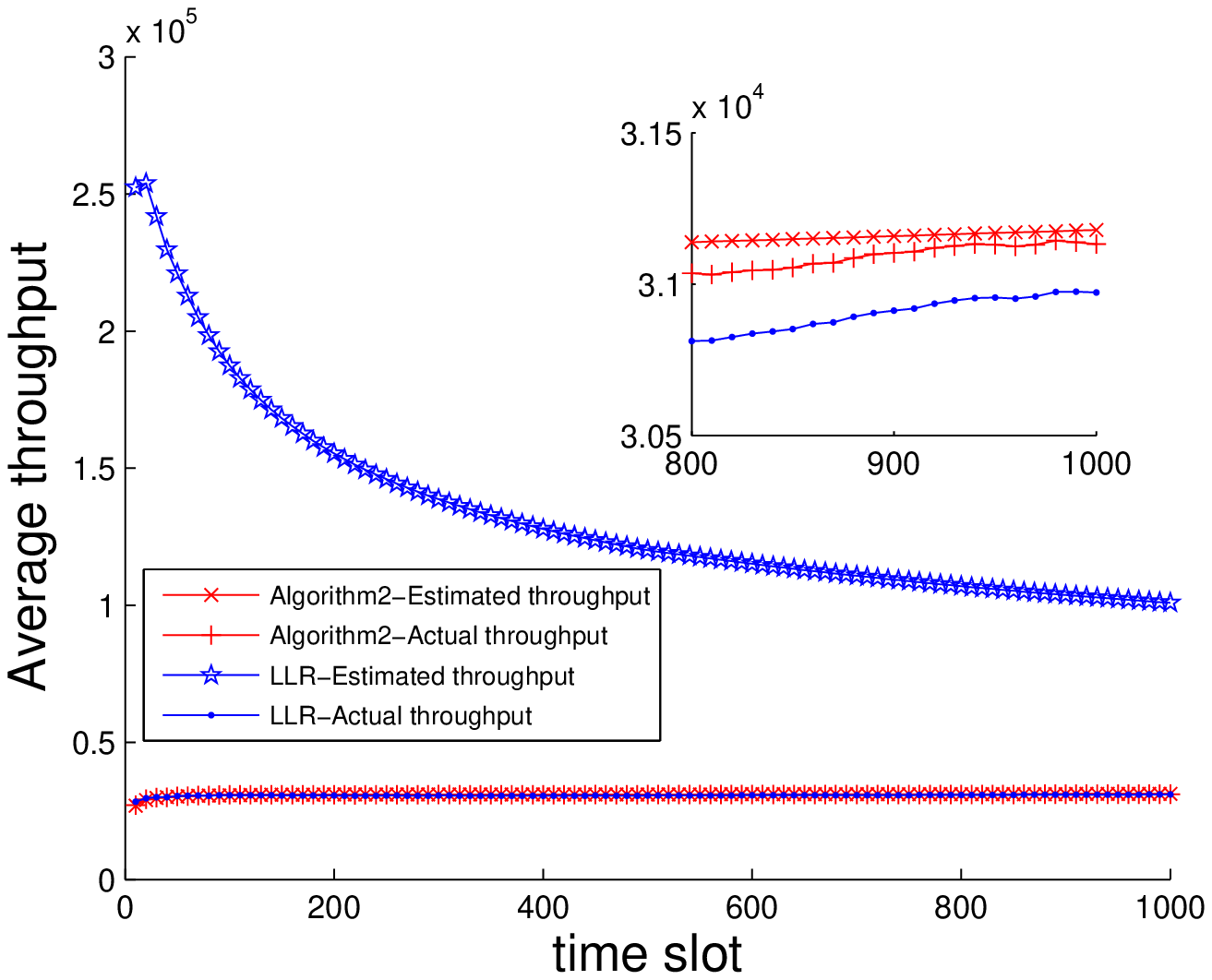}}
  \hspace{0cm}
    \subfigure[5 time slots every period]{
        \label{p5}
        \includegraphics[height=1.2in,width=1.6in]{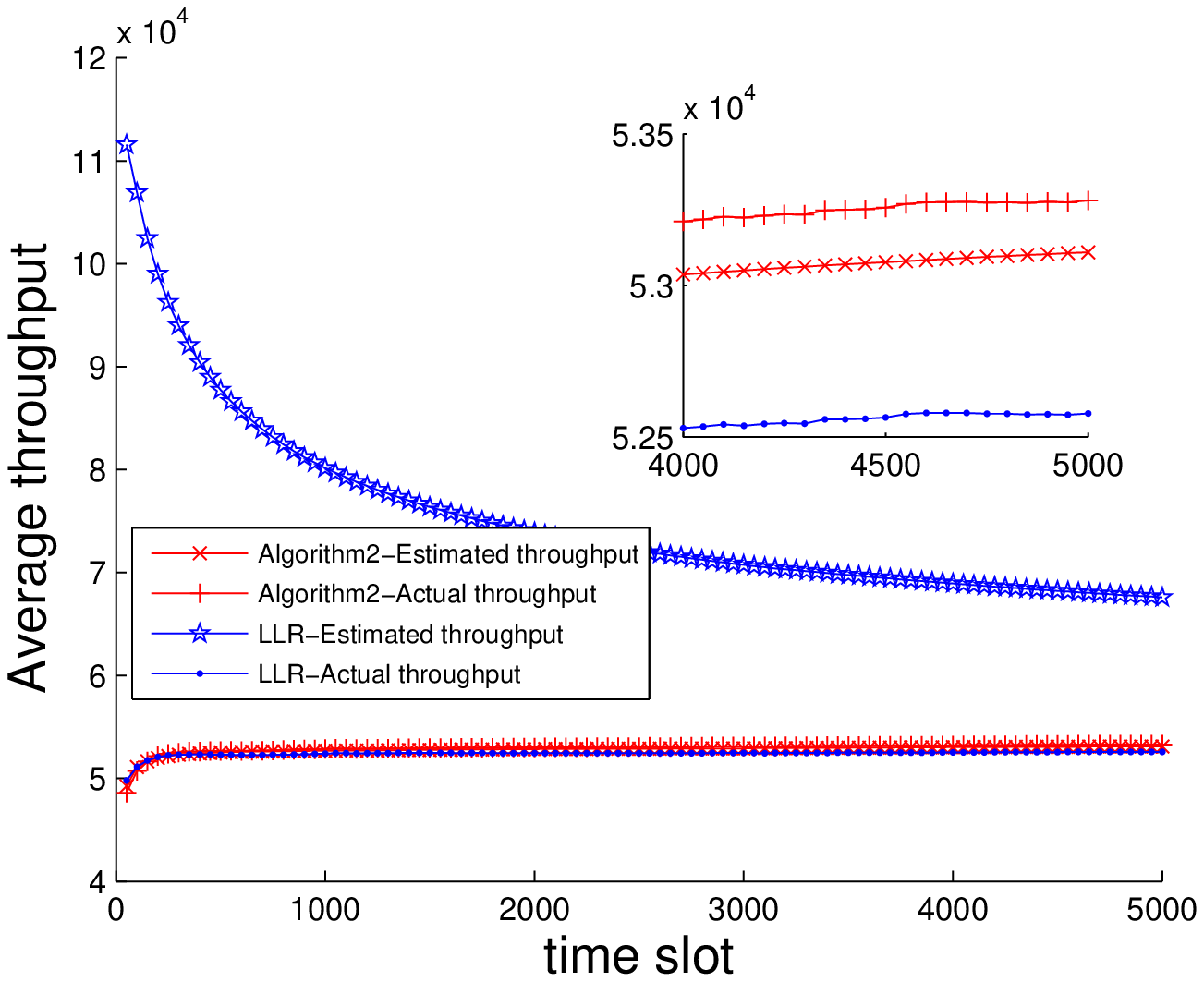}}
  \hspace{0cm}
    \subfigure[10 time slots every period]{
        \label{p10}
        \includegraphics[height=1.2in,width=1.6in]{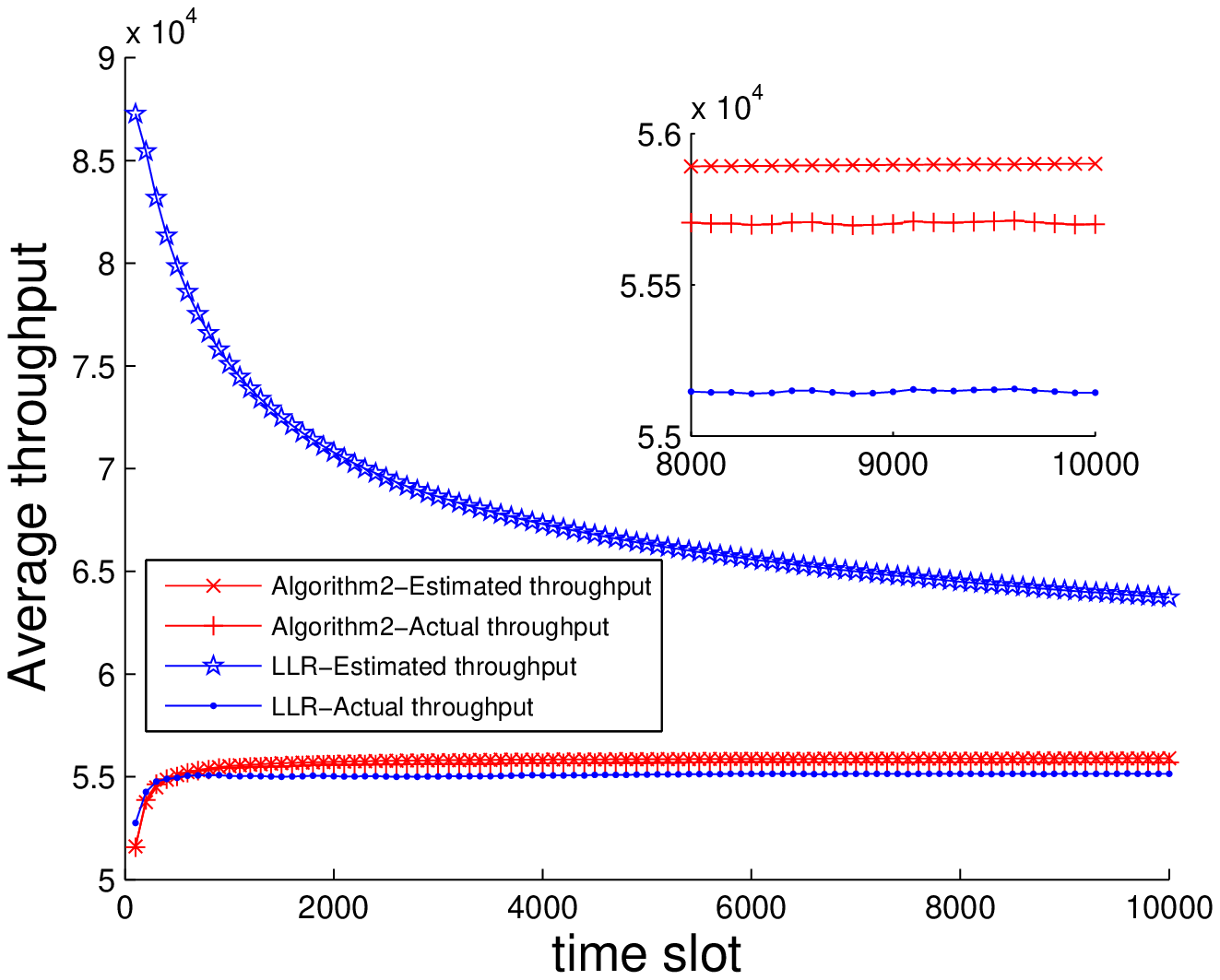}}
  \hspace{0cm}
    \subfigure[20 time slots every period]{
        \label{p20}
        \includegraphics[height=1.2in,width=1.6in]{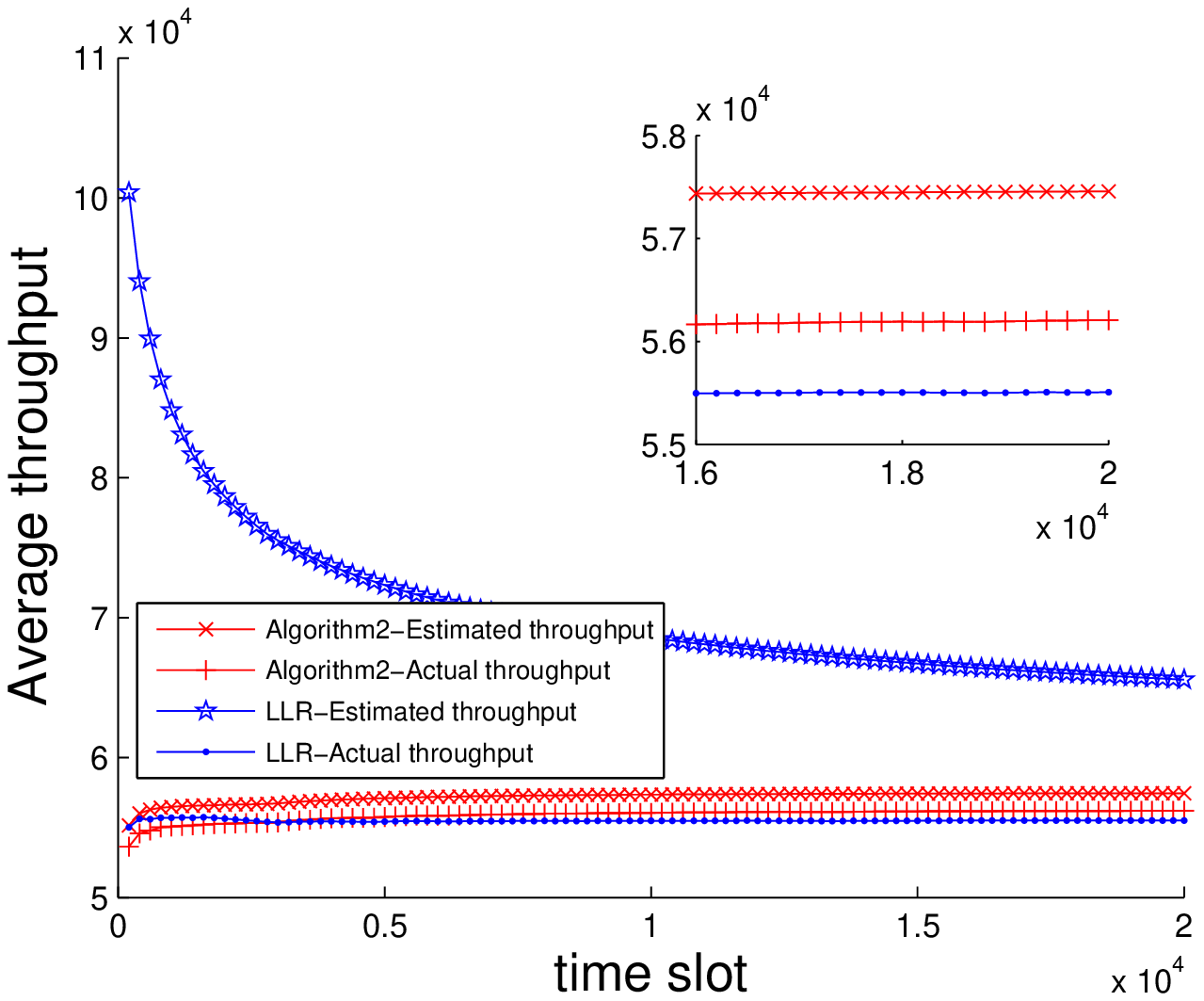}}
   \caption{ Estimated v.s. actual average effective throughput with different period update: comparison with LLR}
 \label{period}
\end{figure*}

  We study the frequent case with $y=1$, and unfrequent cases with stale weight
that is updated periodically with $y= 5, 10, 20$ time slots.
 We conduct each experiment respectively in $1000, 5000,10000,20000$ time slots,
each updating weight $1000$ times.
 The actual effective throughput will be around $1/2, 9/10,19/20,39/40$
of the  ideal throughput without time consuming on strategy decision.
  In Fig.\ref{period}, we can find that the average actual throughput achieved
by both of the algorithms grows to the ideal throughput
as a period lasts more time slots.
    Especially, a significant improvement can be seen between the  frequent case (Fig.\ref{p1})
and the unfrequent case of $y=5$ (Fig.\ref{p5}).
  In the later two  cases, further improvement   is not so obvious
as the proportion of time on learning decreases much more slowly.
  We then compare performance of the two learning policies.
  In each case, we can find that our adopted learning policy is much more accurate
than the LLR learning policy. The difference between the estimated average throughput
and the actual throughput is quite small in our adopted learning policy,
while it is large in the LLR policy. Except the line of estimated throughput by LLR,
difference among other three lines is not obvious in the figures.
    Thus we  show a zoom-in part of the difference on the upper right of each figure.
In these figures, it shows that the actual throughput achieved by our learning policy
is better the LLR policy.  They collaboratively show that unfrequent update has negligible impact
on accuracy of estimation, but significantly improve effective throughput.

%% file: relatedwork.tex
There is  a rich body of results on dynamic spectrum access in cognitive radio networks. As channel availability and quality is unknown to secondary users, they need to conduct a learning process to select good channels. Several literatures address this problem from sequential decision perspective by MAB  approaches, and several from a game theoretic perspective by convergence of equilibrium.

The results using MAB start from single-user play \cite{zhao2008myopic}\cite{ahmad2009optimality}, where each channel evolves as independent and identically distributed  Markov process  with good or bad state. The results are then extended to multi-user play where more than $N > 1$ secondary users select channels among $M$ ones \cite{liu2010distributedmab}, \cite{tekin2012online}, \cite{kalathil2012decentralized}, \cite{liu2013restlessmab}, \cite{anandkumar2010opportunistic}, \cite{anandkumar2011distributed}, \cite{gai2011decentralized}. These works basically assume channel quality  evolving with i.i.d stochastic process over time, and a single-hop network setting where conflict happens if any pair of users choose the same channel simultaneously. For instance, Shu and Krunz \cite{shu2009throughput} propose  a throughput-optimal decision strategy with  stochastic homogeneous channels. This optimal strategy has a threshold structure that indicates whether the channel is good or bad. Anandkumar et al. \cite{anandkumar2011distributed} propose two distributed learning and allocation schemes respectively for the case of pre-allocated ranks for secondary users and non such prior information.

On the other hand, some results consider dynamic spectrum access from an adaptive, game theoretic learning perspective.
M. Maskery, et al.\cite{maskery2009decentralized} model the dynamic channel process as a non-cooperative game for stochastic homogeneous channels, and basically rely on CSMA mechanism to estimate probability of channel contention. In the case of heterogeneous channel quality,
Xu et al. \cite{xu2012opportunistic} construct a potential game to maximize the expected throughput of all secondary users. They implicitly assume a single-hop network case where all users have the same probability to access channels.

We also review the results on network capacity, and related link scheduling problem that maximizes the channel capacity.
There are numerous literatures in this line of work\cite{S:GMS}, \cite{Kodialam2005mobicom}, \cite{jiang2010distributed}, \cite{S:pick2}, originating from the milestone work by Tassiulas et al. \cite{S:MWM1}.
Though both maximizing throughput, the main difference of capacity problems is that they study throughput performance under a known environment without uncertainty of channel quality. The concerned issue is that interference among links constrains the maximum supportable arrival rate at each link that is assumed to have unit capacity. While the problem considered in our work focuses on throughput maximization under unknown and changing link quality, as well as existence of interference. We  need to minimize loss of throughput caused by learning,  as well as time and communication complexity of learning and their impact on throughput performance. 

%% file: conclusion.tex
We proposed an almost throughput optimal channel accessing scheme for multihop cognitive networks. Our scheme  consists of a distributed learning process with low computation and space complexity, and a strategy decision process with low computation and communication complexity. Our distributed implementation does not need extra predefined information on network parameters.

Our works have assumed i.i.d stochastic gain of channels, which is an easy-to-analyze model. Future work will take consideration of adversary case where gains are generated by an adversary that may obliviously or adaptively learn to play against our strategies. Additionally, most works as well as ours minimize\emph{ weak regret} compared to the best static policy, it will be challenging to computation-efficiently  minimize \emph{strong regret} compared to the best dynamic policy.

%% file: channel-access-infocom14-v3.bbl

%% file: channel-access-infocom14-v3.bbl
\begin{thebibliography}{10}
\providecommand{\url}[1]{#1}
\csname url@samestyle\endcsname
\providecommand{\newblock}{\relax}
\providecommand{\bibinfo}[2]{#2}
\providecommand{\BIBentrySTDinterwordspacing}{\spaceskip=0pt\relax}
\providecommand{\BIBentryALTinterwordstretchfactor}{4}
\providecommand{\BIBentryALTinterwordspacing}{\spaceskip=\fontdimen2\font plus
\BIBentryALTinterwordstretchfactor\fontdimen3\font minus
  \fontdimen4\font\relax}
\providecommand{\BIBforeignlanguage}[2]{{%
\expandafter\ifx\csname l@#1\endcsname\relax
\typeout{** WARNING: IEEEtran.bst: No hyphenation pattern has been}%
\typeout{** loaded for the language `#1'. Using the pattern for}%
\typeout{** the default language instead.}%
\else
\language=\csname l@#1\endcsname
\fi
#2}}
\providecommand{\BIBdecl}{\relax}
\BIBdecl

\bibitem{liu2010distributedmab}
K.~Liu and Q.~Zhao, ``Distributed learning in multi-armed bandit with multiple
  players,'' \emph{ IEEE Transactions on Signal Processing}, vol.~58, no.~11,
  pp. 5667--5681, 2010.

\bibitem{tekin2012online}
C.~Tekin and M.~Liu, ``Online learning in decentralized multiuser resource
  sharing problems,'' \emph{arXiv preprint arXiv:1210.5544}, 2012.

\bibitem{kalathil2012decentralized}
D.~Kalathil, N.~Nayyar, and R.~Jain, ``Decentralized learning for multi-player
  multi-armed bandits,'' in \emph{Proc. of  IEEE CDC}, 2012, pp. 3960--3965.

\bibitem{liu2013restlessmab}
H.~Liu, K.~Liu, and Q.~Zhao, ``Learning in a changing world: Restless
  multiarmed bandit with unknown dynamics,'' \emph{ IEEE
  Transactions on Information Theory}, vol.~59, no.~3, pp. 1902--1916, 2013.

\bibitem{anandkumar2010opportunistic}
A.~Anandkumar, N.~Michael, and A.~Tang, ``Opportunistic spectrum access with
  multiple users: learning under competition,'' in \emph{Proc. of  IEEE INFOCOM}, 2010, pp. 1--9.

\bibitem{anandkumar2011distributed}
A.~Anandkumar, N.~Michael, A.~K. Tang, and A.~Swami, ``Distributed algorithms
  for learning and cognitive medium access with logarithmic regret,''
  \emph{IEEE Journal on Selected Areas in Communications}, vol.~29, no.~4, pp.
  731--745, 2011.

\bibitem{gai2011decentralized}
Y.~Gai and B.~Krishnamachari, ``Decentralized online learning algorithms for
  opportunistic spectrum access,'' in \emph{Proc. of  IEEE GLOBECOM}, 2011, pp. 1--6.

\bibitem{lai1985ucb}
T.~L. Lai and H.~Robbins, ``Asymptotically efficient adaptive allocation
  rules,'' \emph{Advances in applied mathematics}, vol.~6, no.~1, pp. 4--22,
  1985.

\bibitem{auer2002finite}
P.~Auer, N.~Cesa-Bianchi, and P.~Fischer, ``Finite-time analysis of the
  multiarmed bandit problem,'' \emph{Machine learning}, vol.~47, no. 2-3, pp.
  235--256, 2002.

\bibitem{agrawal1995sample}
R.~Agrawal, ``Sample mean based index policies with O (log n) regret for the
  multi-armed bandit problem,'' \emph{Advances in Applied Probability}, pp.
  1054--1078, 1995.

\bibitem{gai2012mab}
Y.~Gai, B.~Krishnamachari, and R.~Jain, ``Combinatorial network optimization
  with unknown variables: Multi-armed bandits with linear rewards and
  individual observations,'' \emph{IEEE/ACM Transactions on Networking},
  vol.~20, no.~5, pp. 1466--1478, 2012.

\bibitem{li2012almost}
X.-Y. Li, P.~Yang, Y.~Yan, L.~You, S.~Tang, and Q.~Huang, ``Almost optimal
  accessing of nonstochastic channels in cognitive radio networks,'' in
  \emph{Proc. of  IEEE INFOCOM}, 2012, pp. 2291--2299.

\bibitem{nieberg2005robust}
T.~Nieberg, J.~Hurink, and W.~Kern, ``A robust ptas for maximum weight
  independent sets in unit disk graphs,'' \emph{Graph-theoretic concepts in computer science}, 
  pp. 214--221, 2005.

\bibitem{my2013mab}
Y.~Zhou and X.-Y. Li, ``Multi-armed bandits with combinatorial strategies under
  stochastic bandits,'' \emph{arXiv preprint: http://arxiv.org/abs/1307.5438}, 2013.

\bibitem{mwis2005}
T.~Erlebach, K.~Jansen, and E.~Seidel, ``Polynomial-time approximation schemes
  for geometric intersection graphs,'' \emph{SIAM J. Comput.}, vol.~34, no.~6,
  pp. 1302--1323, Jun. 2005.

\bibitem{intersection2010}
F.~Kammer, T.~Tholey, and H.~Voepel, ``Approximation algorithms for
  intersection graphs,'' pp. 260--273, 2010.

\bibitem{kuhn2005fast}
F.~Kuhn, T.~Moscibroda, T.~Nieberg, and R.~Wattenhofer, ``Fast deterministic
  distributed maximal independent set computation on growth-bounded graphs,''
  in \emph{Distributed Computing}.\hskip 1em plus 0.5em minus 0.4em\relax
  Springer, 2005, pp. 273--287.

\bibitem{huang2008broadcast}
S.-H. Huang, P.-J. Wan, J.~Deng, and Y.~S. Han, ``Broadcast scheduling in
  interference environment,'' \emph{IEEE Transactions on Mobile Computing },
  vol.~7, no.~11, pp. 1338--1348, 2008.

\bibitem{wang2005cds}
Y.~Wang, W.~Wang, and X.-Y. Li, ``Distributed low-cost backbone formation for
  wireless ad hoc networks,'' in \emph{Proc. of ACM MOBIHOC}, 2005, pp. 25--27.

\bibitem{zou2011cds}
F.~Zou, Y.~Wang, X.-H. Xu, X.~Li, H.~Du, P.~Wan, and W.~Wu, ``New
  approximations for minimum-weighted dominating sets and minimum-weighted
  connected dominating sets on unit disk graphs,'' \emph{Theoretical Computer
  Science}, vol. 412, no.~3, pp. 198--208, 2011.

\bibitem{zhao2008myopic}
Q.~Zhao, B.~Krishnamachari, and K.~Liu, ``On myopic sensing for multi-channel
  opportunistic access: Structure, optimality, and performance,''
  \emph{IEEE Transactions on Wireless Communications}, vol.~7, no.~12, pp.
  5431--5440, 2008.

\bibitem{ahmad2009optimality}
S.~Ahmad, M.~Liu, T.~Javidi, Q.~Zhao, and B.~Krishnamachari, ``Optimality of
  myopic sensing in multichannel opportunistic access,'' \emph{IEEE Transactions on 
  Information
  Theory}, vol.~55, no.~9, pp. 4040--4050, 2009.

\bibitem{shu2009throughput}
T.~Shu and M.~Krunz, ``Throughput-efficient sequential channel sensing and
  probing in cognitive radio networks under sensing errors,'' in
  \emph{Proc. of ACM MOBICOM}, 2009,
  pp. 37--48.

\bibitem{maskery2009decentralized}
M.~Maskery, V.~Krishnamurthy, and Q.~Zhao, ``Decentralized dynamic spectrum
  access for cognitive radios: cooperative design of a non-cooperative game,''
  \emph{IEEE Transactions on Communications}, vol.~57, no.~2, pp. 459--469,
  2009.

\bibitem{xu2012opportunistic}
Y.~Xu, J.~Wang, Q.~Wu, A.~Anpalagan, and Y.-D. Yao, ``Opportunistic spectrum
  access in unknown dynamic environment: A game-theoretic stochastic learning
  solution,'' \emph{IEEE Transactions on Wireless Communications}, vol.~11,
  no.~4, pp. 1380--1391, 2012.

\bibitem{S:GMS}
C.~Joo, X.~Lin, and N.~B. Shroff, ``Understanding the capacity region of the
  greedy maximal scheduling algorithm in multi-hop wireless networks,'' in
  \emph{Proc. IEEE INFOCOM}, 2008, pp. 1103--1111.

\bibitem{Kodialam2005mobicom}
M.~Kodialam and T.~Nandagopal, ``Characterizing the capacity region in
  multi-radio multi-channel wireless mesh networks,'' in \emph{Proc. of MOBICOM},
  2005, pp. 73--87.

\bibitem{jiang2010distributed}
L.~Jiang and J.~Walrand, ``A distributed CSMA algorithm for throughput and
  utility maximization in wireless networks,'' \emph{IEEE/ACM Transactions on
  Networking}, vol.~18, no.~3, pp. 960--972, 2010.

\bibitem{S:pick2}
S.-J. Tang, X.-Y. Li, X.~Wu, Y.~Wu, X.~Mao, P.~Xu, and G.~Chen, ``Low
  complexity stable link scheduling for maximizing throughput in wireless
  networks,'' in \emph{Proc. of  IEEE  SECON}, 2009, pp. 1--9.

\bibitem{S:MWM1}
L.~Tassiulas and A.~Ephremides, ``Stability properties of constrained queueing
  systems and scheduling policies for maximum throughput in multihop radio
  networks,'' \emph{IEEE/ACM Transactions on Automatic Control}, vol. $37$, pp.
  1936--1948, 1992.

\end{thebibliography}
